\documentclass[pdflatex,sn-mathphys-num]{sn-jnl}

\usepackage{geometry}
\AtBeginDocument{%
  \newgeometry{a4paper,margin=2cm}%
}
\usepackage{graphicx}%
\usepackage{multirow}%
\usepackage{amsmath,amssymb,amsfonts}%
\usepackage{amsthm}%
\usepackage{mathrsfs}%
\usepackage{xcolor}%
\usepackage{textcomp}%
\usepackage{manyfoot}%
\usepackage{footmisc} 
\usepackage{algorithm}%
\usepackage{algorithmicx}%
\usepackage{algpseudocode}%
\usepackage{listings}%
\usepackage{color}%
\usepackage[normalem]{ulem}

\theoremstyle{thmstyleone}%

\theoremstyle{thmstyletwo}%
\theoremstyle{thmstylethree}%
\begin{document}

\title[Article Title]{Reactive near-field subwavelength microwave imaging with a non-invasive Rydberg probe}


\author[1,2,†]{\fnm{Chaoyang} \sur{Hu}}
\author*[1,2,†,]{\fnm{Mingyong} \sur{Jing}}\email{jmy@sxu.edu.cn}
\author[1,2,]{\fnm{Zongkai} \sur{Liu}}
\equalcont{These authors contributed equally to this work.}

\author[1,2]{\fnm{Shaoxin} \sur{Yuan}}
\author[1,2]{\fnm{Bin} \sur{Wu}}
\author[1,2]{\fnm{Yan} \sur{Peng}}
\author[1,2]{\fnm{Tingting} \sur{Li}}

\author[1,2]{\fnm{Wenguang} \sur{Yang}}
\author[1,2]{\fnm{Junyao} \sur{Xie}}
\author[1,2]{\fnm{Hao} \sur{Zhang}}

\author[1,2]{\fnm{Liantuan} \sur{Xiao}}
\author[1,2]{\fnm{Suotang} \sur{Jia}}
\author*[1,2,]{\fnm{Linjie} \sur{Zhang}}\email{zlj@sxu.edu.cn}

\affil[1]{\orgdiv{State Key Laboratory of Quantum Optics Technologies and Devices}, \orgname{Institute of Laser Spectroscopy}, \orgaddress{\city{Taiyuan}, \postcode{030006}, \state{Shanxi}, \country{China}}}

\affil[2]{\orgdiv{Collaborative Innovation Center of Extreme Optics}, \orgname{Shanxi University}, \orgaddress{\city{Taiyuan}, \postcode{030006}, \state{Shanxi}, \country{China}}}


\abstract{Non-invasive microwave field imaging—accurately mapping field distributions without perturbing them—is essential in areas such as aerospace engineering, biomedical imaging and integrated-circuit diagnostics. Conventional metal probes, however, inevitably perturb reactive near fields: they act as strong scatterers that drive induced currents and secondary radiation, remap evanescent components and thereby degrade both accuracy and spatial resolution, particularly in the reactive near-field regime that is most relevant to these applications. Here we demonstrate, to our knowledge for the first time, reactive near-field subwavelength imaging of microwave fields using the quantum non-demolition properties of Rydberg atoms, realized with a compact, non-invasive single-ended fibre-integrated Rydberg probe engineered to minimize field disturbance. The probe achieves an imaging resolution of {\unboldmath$\lambda/56$}, and the measured field distributions agree with full-wave simulations with structural similarity approaching unity, confirming both its subwavelength spatial resolution and its genuinely non-invasive character compared with conventional metal-based probes. Because the atomic sensor is intrinsically isotropic, the same device can faithfully image multi-dimensional field structures without orientation-dependent calibration. Our results therefore establish a general, non-invasive route to high-accuracy, subwavelength reactive near-field microwave imaging, with particular promise for applications such as chip-defect detection and integrated-circuit diagnostics, where even small perturbations by the probe can mask the underlying physics of interest.} 


\maketitle

\section{Introduction}\label{sec1}

Reactive field components and evanescent waves in the microwave reactive near field (RNF) make it possible to circumvent the diffraction limit that constrains far-field imaging, thereby providing access to rich subwavelength structural information~\cite{RN1,RN2,RN3}. As a consequence, microwave RNF imaging can deliver high-resolution visualization of internal features in both engineered materials and biological samples. This capability underpins the broad use of RNF techniques across diverse domains, including the aerospace industry, advanced manufacturing~\cite{RN4,RN5,RN6,RN7}, civil infrastructure~\cite{RN8,RN9,RN10,RN11,RN12,RN13}, biomedical diagnostics~\cite{RN14,RN15,RN16,RN17,RN18} and integrated-circuit (IC) testing and failure analysis~\cite{RN19,RN20,RN21,RN22,RN23,RN24,RN25,RN26,RN27}.

Despite this potential, accurate and truly non-invasive microwave RNF imaging remains challenging. Classical metallic probes—such as open-ended coaxial tips, waveguides and dipoles~\cite{RN28,RN29,RN30,RN31,RN32,RN33}—are fundamentally limited by their conductivity. By imposing a conducting boundary, a metal probe draws large induced currents that re-radiate and remap the native evanescent field, so that the measurement primarily reflects the field disturbance created by the probe rather than the undisturbed field itself~\cite{RN34,RN35,RN36}. The resulting perturbation is determined by the probe radiation characteristics and depends sensitively on its orientation and position relative to the source~\cite{RN37,RN38,RN39}, leading to complex electromagnetic coupling between probe and field that undermines accuracy and repeatability. Probe-compensation algorithms can, in favourable cases, enable post-processed imaging in the radiating near field (RDNF), but the rich evanescent content and strong coupling found in the RNF make such compensation increasingly difficult and often unreliable~\cite{RN40}.

To mitigate probe invasiveness, alternative sensing platforms have been explored, including electro-optic probes~\cite{RN41,RN42,RN43}, diamond nitrogen-vacancy (NV) centres~\cite{RN44,RN45} and other quantum-sensing architectures. Among these, microwave-field sensing based on Rydberg atoms~\cite{RN46,RN47,RN48} has emerged as a particularly powerful approach for near-field imaging~\cite{RN49,RN50,RN51}, owing to its intrinsic SI-traceability and the fact that the atoms themselves can be made almost electromagnetically transparent to the field being measured~\cite{RN52,RN53}.

In this work, we introduce a non-invasive, single-ended fibre-integrated Rydberg probe that fundamentally differs from previously reported fibre-integrated designs~\cite{RN54,RN55,RN56,RN57,RN58}. Combined with a robotic-arm-based three-dimensional imaging platform and a dedicated optical detection and reconstruction algorithm, this probe enables subwavelength ($\lambda/56$), multi-dimensional RNF imaging of microwave fields. The probe structure is engineered entirely from non-metallic materials to suppress field perturbations normally introduced by Rydberg-probe integration. Together with the intrinsically isotropic response of the atomic sensor~\cite{RN59}, this design yields measured images whose structural similarity (SSIM) to full-wave simulations approaches unity. Direct comparison with a classical compact antenna of similar size, operated under identical conditions, clearly demonstrates the superior imaging fidelity of the atomic probe. Furthermore, because the probe perturbs the field only weakly, it allows direct characterization of field distortions caused by additional external objects placed in the RNF. Overall, this work establishes a practical route toward applying Rydberg-atom-based quantum technology to RNF microwave subwavelength imaging and, in particular, to non-invasive diagnostics of microwave components and integrated circuits.

\section{Results}\label{sec2}

\subsection{Experimental setup for microwave RNF imaging}\label{subsec1}

\begin{figure}[!htbp]
\centering
\includegraphics[width=1\textwidth]{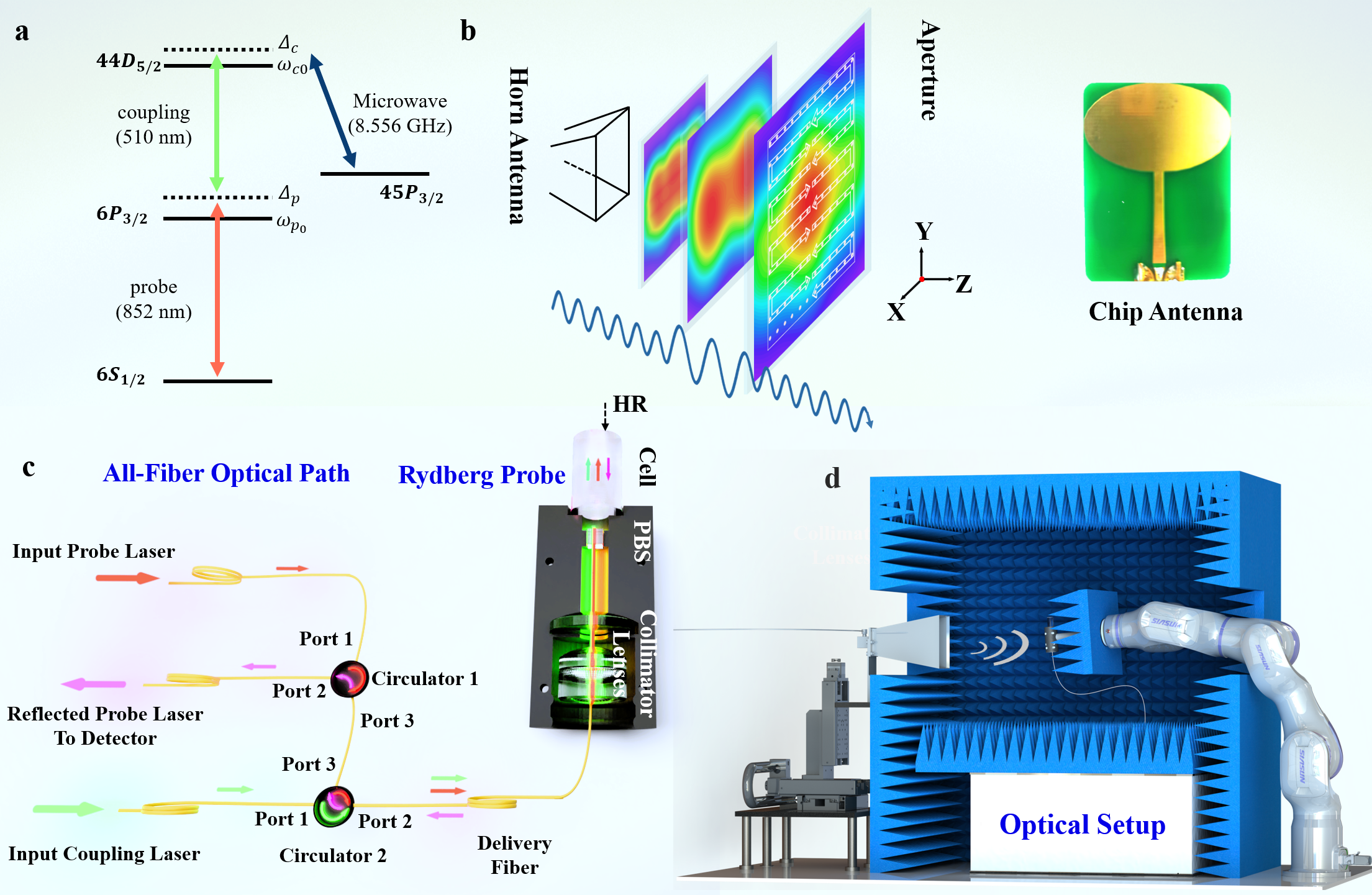}
\caption{\textbf{Schematic of the RNF imaging platform.} \textbf{a} Cesium level scheme and optical geometry. The 852~nm probe drives $|6S_{1/2}\rangle \rightarrow |6P_{3/2}\rangle$ with detuning $\Delta_{p}$; the 510~nm coupling drives $|6P_{3/2}\rangle \rightarrow |44D_{5/2}\rangle$ with detuning $\Delta_{c}$; a microwave at 8.556~GHz couples $|44D_{5/2}\rangle \rightarrow |45P_{3/2}\rangle$. \textbf{b} Antennas under test (AUT) and scan geometry. A standard horn (aperture $138~\mathrm{mm}\times 107~\mathrm{mm}$) and a UWB chip antenna (elliptical patch, major axis $19~\mathrm{mm}$, minor axis $9~\mathrm{mm}$) are driven; the near-field scan trajectory and Cartesian axes are indicated. \textbf{c} Non-invasive, single-ended fibre-integrated Rydberg probe. The 852~nm probe (orange) and 510~nm coupling (green) are delivered via two cascaded multimode circulators to the probe head. An 852~nm high-reflectivity coating (HR) on the inner surface of the cell’s rear window reflects the probe to generate a counter-propagating beam (magenta), which returns via the PBS, collimators and circulators—thereby enabling both co- and counter-propagating configurations with the coupling beam. \textbf{d} Robotic-arm three-dimensional imaging platform. The AUT and probe head operate inside a microwave-absorbing enclosure, while the optical setup remains outside.}\label{fig1}
\end{figure}

Fig.~\ref{fig1} summarizes the architecture of the non-invasive RNF imaging platform based on Rydberg atoms. The system comprises three main subsystems: (i) an RF excitation chain, (ii) a non-invasive single-ended fibre-integrated Rydberg probe and (iii) a robotic-arm-based three-dimensional positioning and imaging stage. The relevant cesium level structure is shown in Fig.~\ref{fig1}a: two optical fields at 852~nm (probe) and 510~nm (coupling) form a ladder scheme, while a centimetre-wave field at 8.556~GHz couples two Rydberg states and induces an Autler--Townes (AT) splitting.

The RF subsystem consists of a microwave source and the antennas under test (AUTs), shown schematically in Fig.~\ref{fig1}b: a standard gain horn and an ultra-wideband (UWB) chip antenna, both driven at 8.556~GHz. For the horn antenna, Cartesian axes are indicated, and the coordinate origin is defined at a position 53.5~mm along the negative $\mathrm{X}$-direction and 69~mm along the negative $\mathrm{Y}$-direction from the aperture centre, which is used consistently throughout the imaging experiments.

A key element of the platform is the single-ended, fibre-integrated Rydberg probe, designed to be entirely free of metal near the sensing volume so as to minimize field perturbation. As illustrated in Fig.~\ref{fig1}c, the 852~nm probe beam is routed from port~1$\rightarrow$2 of circulator~1 and then 3$\rightarrow$2 of circulator~2 into the probe head, while the 510~nm coupling beam is injected from port~1$\rightarrow$2 of circulator~2 into the same path. Inside the probe, the co-propagating beams pass through collimators and a polarizing beam splitter (PBS) before entering a cesium vapour cell whose back face is coated with a $0^\circ$ high-reflectivity (HR) mirror at 852~nm. The probe beam reflects from this HR surface to generate a counter-propagating component, which exits via the PBS and collimators, retraces the fibre path through circulator~2 (2$\rightarrow$3) and circulator~1 (2$\rightarrow$3) and is finally detected on a photodiode.

All-optical readout of the microwave field relies on electromagnetically induced transparency (EIT) and AT splitting of the Rydberg resonance. In a thermal vapour, Doppler mismatch between the probe and coupling beams broadens the EIT feature~\cite{RN64}; this broadening is significantly stronger for co-propagating beams than for counter-propagating ones. To exploit the narrowest resonance, we implement a velocity-selective scheme in which the 852~nm probe is detuned by $\Delta_{p}=+80$~MHz from the zero-velocity transition, which spectrally separates the co- and counter-propagating EIT contributions (see Supplementary Information Section S.1). The local microwave-field amplitude is then extracted from the AT splitting of the counter-propagating EIT peak~\cite{RN51}, which is least affected by Doppler mismatch and thus offers the highest spectral resolution.

The robotic-arm imaging subsystem (Fig.~\ref{fig1}d) integrates a robotic arm with auxiliary translation stages to position the probe in three dimensions relative to the AUT. While the arm provides dexterous motion and orientation control, the stages enlarge the reachable workspace. Owing to their finite travel ranges, neither component alone can span the entire imaging volume of interest. We therefore coordinate both actuators via a motion-planning algorithm that constrains the relative pose of the AUT and probe to follow a prescribed scan trajectory, such as that indicated in Fig.~\ref{fig1}b. At each scan point, the system automatically records the probe spectrum and the corresponding three-dimensional position. These data are subsequently processed to extract the local field amplitude $E(\mathrm{X},\mathrm{Y},\mathrm{Z})$ and reconstruct the microwave RNF image.

\subsection{Microwave non-invasive imaging in the RNF of a horn antenna}\label{subsec2}

Non-invasive near-field metrology is central to modern antenna calibration: by measuring both amplitude and phase across the aperture, one can reconstruct the radiated wavefront and realize compact test ranges with reduced sensitivity to multipath reflections. Classical metallic probes, however, couple strongly to the AUT, perturbing the field and often forcing measurements to be performed in the Fresnel region, where probe-compensation procedures are applicable but tend to break down in the RNF~\cite{RN36}. In this section, we show that the Rydberg-atom probe provides minimally perturbative readout, recovering the near-field distribution without any compensation—both in the RDNF and, crucially, in the RNF—and thereby enabling genuinely non-invasive antenna calibration.

\begin{figure}[!htbp]
\centering
\includegraphics[width=1\textwidth]{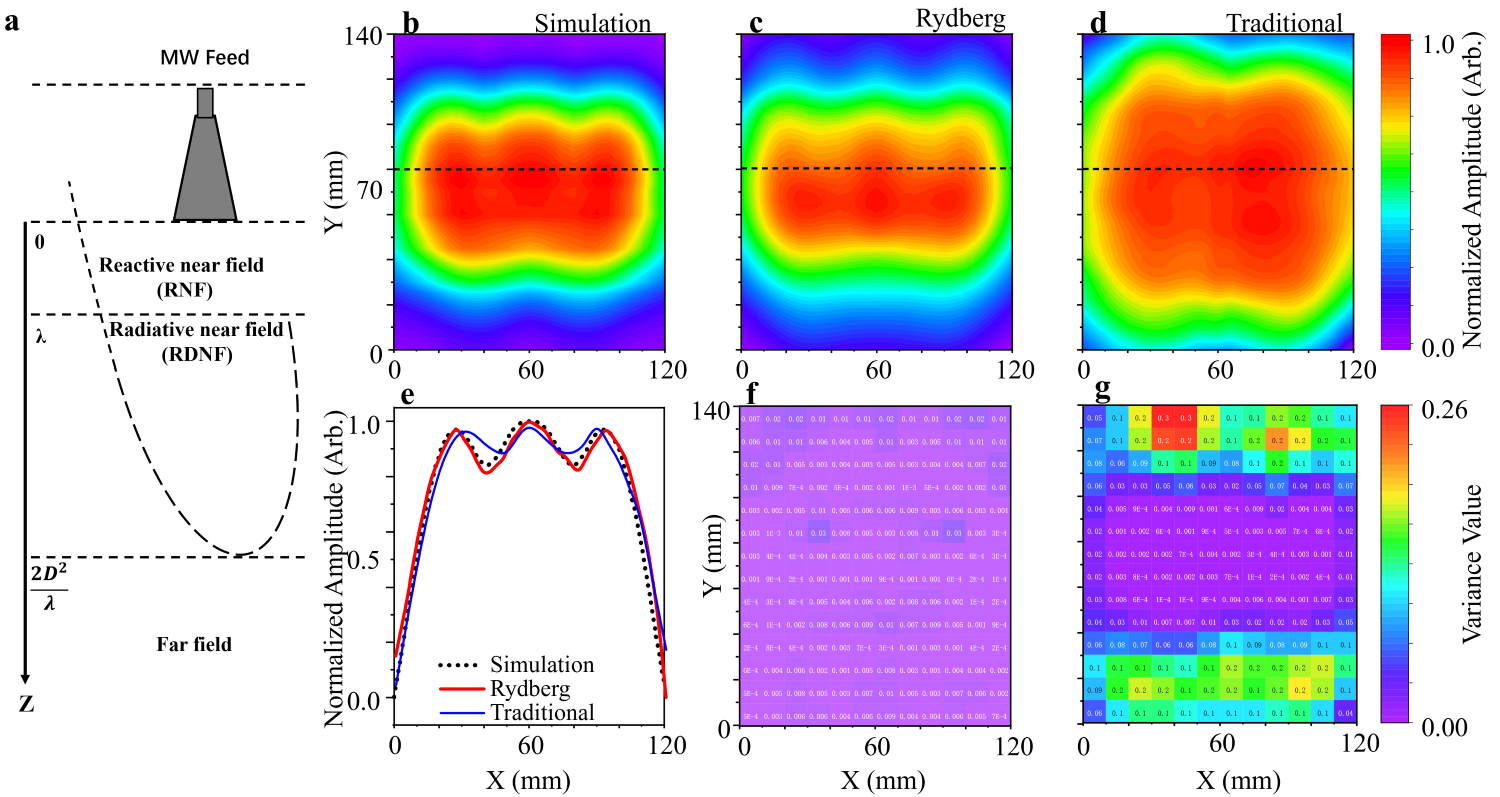}
\caption{\textbf{Microwave field mapping in the RNF of a standard-gain horn.}
\textbf{a} Schematic bounds of the reactive near field, radiating near field and far field for a rectangular horn.
\textbf{b} Theoretical (CST-simulated) field distribution at the $\mathrm{Z}=17.5~\mathrm{mm}$ plane, used as the undisturbed-field reference.
\textbf{c},\textbf{d} Measured two-dimensional field-strength maps obtained with the Rydberg probe and, for comparison, a compact omnidirectional metal antenna with an elliptical aperture comparable to that of the cesium vapour cell (Fig.~\ref{fig1}b) ($\mathrm{XY}$ plane at $\mathrm{Z}=17.5~\mathrm{mm}$; step sizes $\Delta \mathrm{X}=\Delta \mathrm{Y}=1~\mathrm{mm}$).
\textbf{e} One-dimensional profiles along $\mathrm{X}$ at $\mathrm{Y}=80~\mathrm{mm}$ (step $\Delta \mathrm{X}=2~\mathrm{mm}$).
\textbf{f},\textbf{g} Pointwise difference maps relative to the simulated reference for the Rydberg probe and the metal antenna, respectively.}\label{fig2}
\end{figure}

As shown in Fig.~\ref{fig2}a, the space around a transmitting antenna is partitioned into the reactive near field, the radiating near field and the far field. Following Ref.~\cite{RN62}, for a rectangular horn with maximum linear aperture $D$, the axial distance $\mathrm{Z}$ from the horn aperture falls in the reactive or radiating near-field regions when
\begin{equation}
\label{deqn_eq_1}
0<\mathrm{Z}<\lambda,\quad \lambda<\mathrm{Z}<\frac{2D^2}{\lambda},
\end{equation}
where $\lambda$ is the wavelength. For the horn used in our experiment (aperture $138\,\mathrm{mm}\times 107\,\mathrm{mm}$, giving $D=175\,\mathrm{mm}$) operated at $8.556\,\mathrm{GHz}$ (so $\lambda \approx 35\,\mathrm{mm}$), these relations yield an RNF spanning $0<\mathrm{Z}<\lambda \approx 35\,\mathrm{mm}$ and an RDNF spanning $\lambda<\mathrm{Z}<2D^2/\lambda \approx 1.5\,\mathrm{m}$ (about $43\,\lambda$), as indicated in Fig.~\ref{fig2}a. 

To quantify non-invasive imaging performance in the RNF, we use the Rydberg probe of Fig.~\ref{fig1}c to scan an $\mathrm{XY}$ plane located half a wavelength from the horn aperture ($\mathrm{Z}=17.5~\mathrm{mm}$) over an area of $120~\mathrm{mm}\times 140~\mathrm{mm}$. The corresponding theoretical field map in Fig.~\ref{fig2}b is obtained using CST Studio Suite with the vendor-provided horn model and serves as an undisturbed-field reference.

The image acquired with the Rydberg sensor (Fig.~\ref{fig2}c) reproduces the main and side lobes of the horn pattern with high fidelity and closely matches the simulation. The distribution is nearly symmetric along $\mathrm{X}$, while a slight asymmetry along $\mathrm{Y}$ is consistent with a $\sim 5^{\circ}$ downward tilt of the horn mount, as verified by CST simulations with the same tilt. For a fair comparison, we also use a compact omnidirectional metal antenna whose elliptical aperture matches that of the cesium vapour cell (Fig.~\ref{fig1}b), so that any differences arise from probe invasiveness rather than aperture size. As shown in Fig.~\ref{fig2}d, this metal antenna significantly perturbs the RNF and distorts the native lobe structure.

A more detailed comparison is provided in Fig.~\ref{fig2}e, which overlays one-dimensional profiles at $\mathrm{Y}=80~\mathrm{mm}$ taken from Figs.~\ref{fig2}b–d. The Rydberg-trace profile tracks the simulated reference closely, whereas the metal-antenna profile exhibits pronounced deviations, underscoring the invasive nature of the metal probe. Quantitatively, Figs.~\ref{fig2}f and \ref{fig2}g show pointwise difference maps relative to the reference; the error is substantially smaller for the Rydberg probe across the entire field of view.

Structural fidelity between field maps is quantified using the structural similarity (SSIM) index~\cite{RN63}, with values close to unity indicating high image similarity:
\begin{equation}
\label{deqn_eq_2}
\mathrm{SSIM}(i,j)=\frac{(2\mu_i\mu_j+C_1)(2\sigma_{ij}+C_2)}{(\mu_i^2+\mu_j^2+C_1)(\sigma_i^2+\sigma_j^2+C_2)}.
\end{equation}
Here, $\mu_i$ and $\mu_j$ denote the local means, $\sigma_i$ and $\sigma_j$ the local standard deviations and $\sigma_{ij}$ the local covariance of the normalized electric-field-strength values within each sliding window for images $i$ and $j$, respectively. The constants $C_1=(K_1 L)^2$ and $C_2=(K_2 L)^2$ are small positive terms introduced for numerical stability, where $L$ is the dynamic range of the normalized field values ($L=1$ here), and we use the standard choices $K_1=0.01$ and $K_2=0.03$. Treating the field strength at each $(\mathrm{X},\mathrm{Y})$ coordinate as an image pixel, we obtain $\mathrm{SSIM}(b,d)=0.773$ for the non-invasive reference (b) versus the metal-antenna map (d), and $\mathrm{SSIM}(b,c)=0.971$ for the reference (b) versus the Rydberg map (c), demonstrating that the Rydberg measurement faithfully reproduces the undisturbed field distribution. A more detailed derivation of Eq.~\eqref{deqn_eq_2} is given in Supplementary Information Section S.2.

Taken together, the one-dimensional profiles, difference maps and SSIM metrics consistently show that the Rydberg probe provides minimally perturbative, high-fidelity imaging of the RNF, whereas the metal antenna substantially distorts the native field. The same non-invasive performance is also observed in the compact near field of an ultra-wideband chip antenna, where the Rydberg probe accurately recovers the elliptical radiation pattern while a metal probe fails (Supplementary Information Section S.3).

\begin{figure}[!htbp]
\centering
\includegraphics[width=1\textwidth]{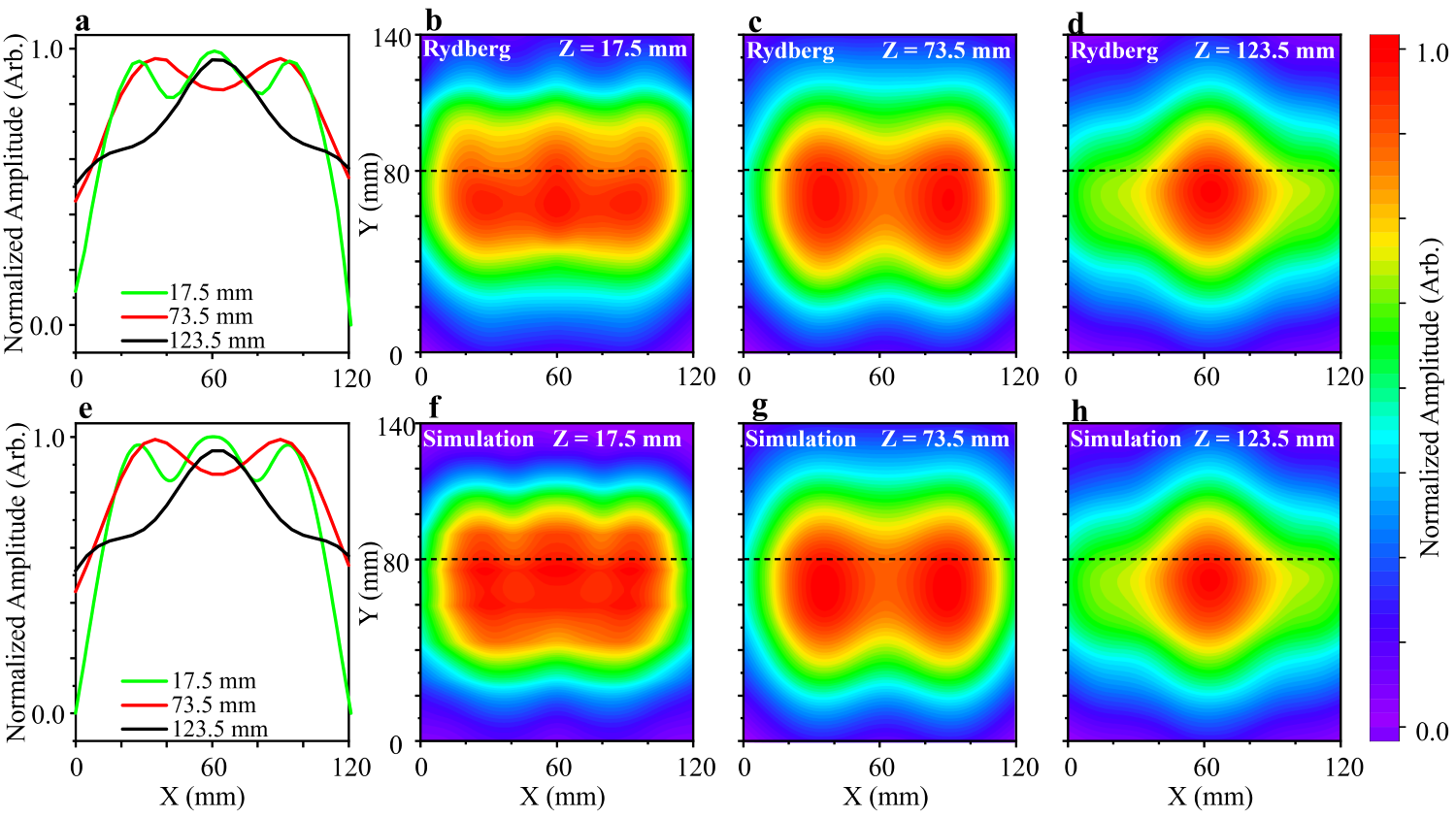}
\caption{\textbf{Microwave near--field characterization and RNF--RDNF evolution of a horn antenna.}
\textbf{a} One-dimensional electric-field profiles along $\mathrm{X}$ at $\mathrm{Y}=80~\mathrm{mm}$ extracted from the Rydberg-probe measurements in \textbf{b}–\textbf{d}.
\textbf{b}–\textbf{d} Measured electric-field maps obtained with the Rydberg probe at $\mathrm{Z}=17.5~\mathrm{mm}$ (RNF) and at $\mathrm{Z}=73.5~\mathrm{mm}$ and $\mathrm{Z}=123.5~\mathrm{mm}$ (RDNF).
\textbf{e}–\textbf{h} Corresponding CST-simulated field maps serving as undisturbed-field references for panels \textbf{a}–\textbf{d}.}
\label{fig3}
\end{figure}

Beyond accurate mapping, the resulting high-quality amplitude data form a robust basis for antenna diagnostics, including stable phase retrieval via near--field--to--far--field (NF--FF) transformation~\cite{RN64} with improved resilience to multipath and stray electromagnetic noise. To examine the evolution from RNF to RDNF in more detail, we measure the microwave electric-field distribution with the Rydberg probe on three $\mathrm{XY}$ planes along the horn axis, at $\mathrm{Z}=17.5~\mathrm{mm}$ (RNF) and at $\mathrm{Z}=73.5~\mathrm{mm}$ and $\mathrm{Z}=123.5~\mathrm{mm}$ (RDNF), as shown in Figs.~\ref{fig3}b–d. The corresponding CST reference maps are plotted in Figs.~\ref{fig3}e–h. In the RNF, the main lobe and two side lobes are clearly resolved. As the field propagates into the RDNF, path-length differences and diverging propagation directions cause these lobes to merge—first into two broader features and eventually into a single composite profile.

Structural-similarity analysis yields $\mathrm{SSIM}(b,f)=0.971$, $\mathrm{SSIM}(c,g)=0.998$ and $\mathrm{SSIM}(d,h)=0.998$, demonstrating excellent agreement between measurement and reference across all planes, with near-perfect correspondence in the RDNF. The minimal perturbation introduced by the probe thus supports reliable NF--FF transformation from RDNF data and enables accurate antenna characterization and far-field reconstruction.

\subsection{Non-invasive imaging of subwavelength targets in the microwave reactive near field}\label{subsec4}

The previous sections focused on antenna RNF distributions in the absence of additional scatterers. We now consider a more demanding scenario in which a subwavelength object resides within the RNF and must be imaged via its perturbation of the field. The central challenge is to resolve the object’s scattering signature without the probe itself introducing a comparable or larger distortion. The strongly non-invasive character of the Rydberg probe is therefore crucial, as it ensures that the measured field perturbation originates primarily from the target rather than from the sensor.

\begin{figure}[h]
\centering
\includegraphics[width=0.9\textwidth]{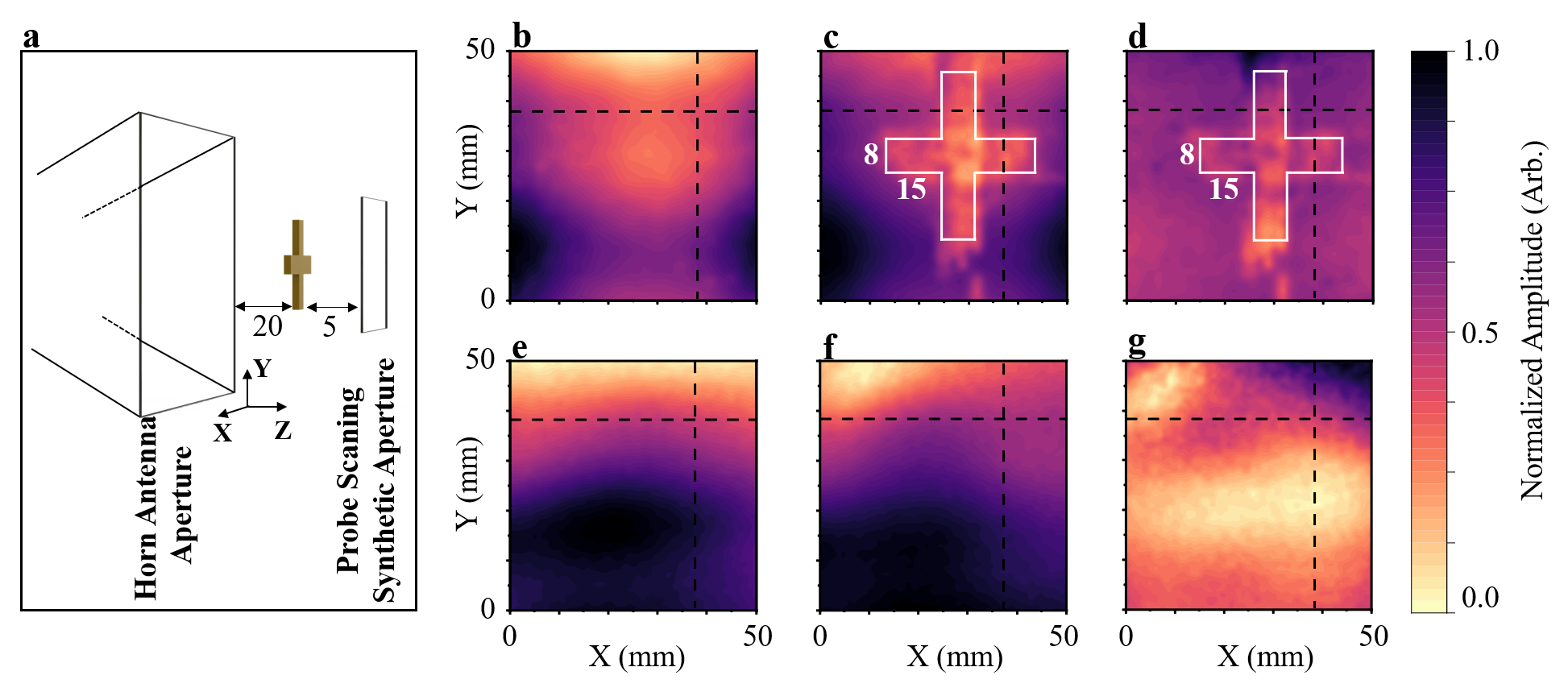}
\caption{\textbf{Non-invasive imaging of a subwavelength target in the microwave reactive near field.}
\textbf{a} Schematic of the target-imaging configuration in the reactive near field.
\textbf{b}–\textbf{d} Rydberg-atom images: background without the target (\textbf{b}), image with the target present (\textbf{c}) and background-subtracted image (\textbf{d}); step sizes $\Delta \mathrm{X} = 0.1~\mathrm{mm}$ and $\Delta \mathrm{Y} = 0.1~\mathrm{mm}$.
\textbf{e}–\textbf{g} Corresponding images obtained with the traditional metal probe (Fig.~\ref{fig1}b): background (\textbf{e}), target present (\textbf{f}) and background-subtracted image (\textbf{g}).}
\label{fig4}
\end{figure}

As a representative subwavelength target, we use a patterned copper tag (dimensions in Fig.~\ref{fig4}c and d). The tag is placed at the centre position ($\mathrm{X}=60~\mathrm{mm}$, $\mathrm{Y}=70~\mathrm{mm}$) of a standard-gain horn antenna radiating at $8.556~\mathrm{GHz}$, approximately $20~\mathrm{mm}$ from the horn aperture. The two-dimensional microwave electric-field distribution is measured on a plane $5~\mathrm{mm}$ in front of the tag using the Rydberg probe, scanning a $50~\mathrm{mm} \times 50~\mathrm{mm}$ area. Fig.~\ref{fig4}b shows the background field distribution without the target, and Fig.~\ref{fig4}c shows the field with the target present. The tag blocks and redistributes the microwave field, and because diffraction over the short propagation distance ($\sim 5~\mathrm{mm}$) is limited, a clear outline of the target appears in the measured field.

To suppress the intrinsic inhomogeneity of the RNF background, we subtract the background image in Fig.~\ref{fig4}b from Fig.~\ref{fig4}c, obtaining the differential image in Fig.~\ref{fig4}d in which the background is substantially flattened and the target contour is further enhanced. For comparison, we repeat the same measurement using a traditional metal chip antenna (Fig.~\ref{fig1}b) as the probe. The corresponding images are shown in Figs.~\ref{fig4}e–g.

\begin{figure}[h]
\centering
\includegraphics[width=0.9\textwidth]{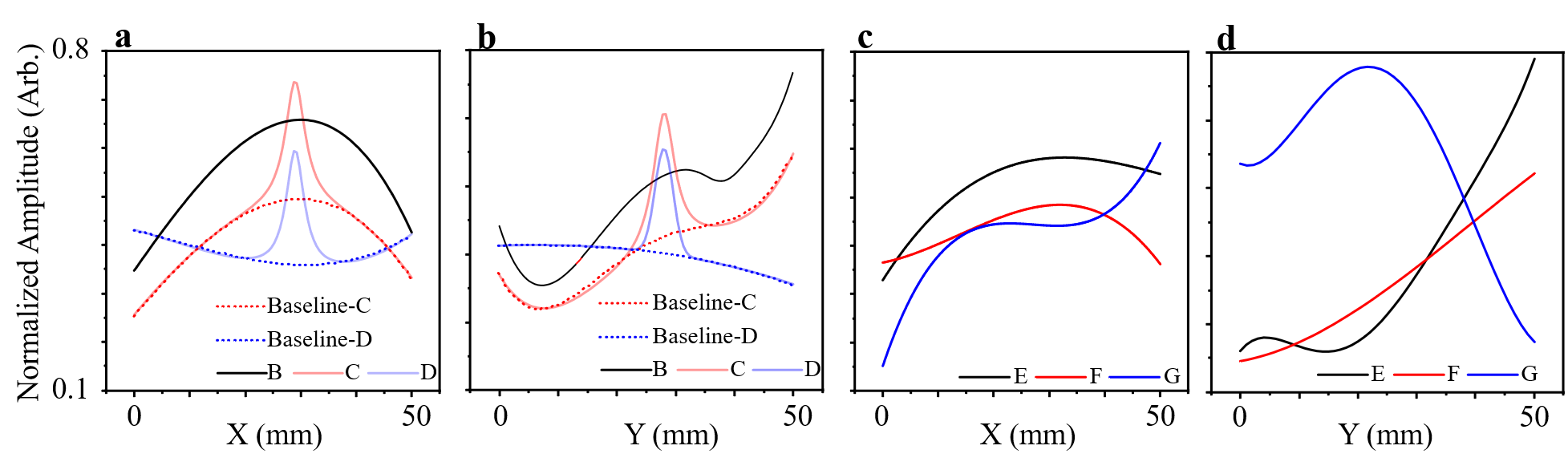}
\caption{\textbf{Imaging performance metrics.}
\textbf{a} One-dimensional microwave electric-field profiles at $\mathrm{Y}=38~\mathrm{mm}$ extracted from Fig.~\ref{fig4}b–d (Rydberg probe).
\textbf{b} One-dimensional profiles at $\mathrm{X}=38~\mathrm{mm}$ from Fig.~\ref{fig4}b–d.
\textbf{c} One-dimensional profiles at $\mathrm{Y}=38~\mathrm{mm}$ from Fig.~\ref{fig4}e–g (metal probe).
\textbf{d} One-dimensional profiles at $\mathrm{X}=38~\mathrm{mm}$ from Fig.~\ref{fig4}e–g.}
\label{fig5}
\end{figure}

To quantify the imaging performance, we extract one-dimensional profiles of the microwave electric field at $\mathrm{Y}=38~\mathrm{mm}$ from Fig.~\ref{fig4}b–d, yielding curves B, C and D in Fig.~\ref{fig5}a. The dashed curves (Baseline-C and Baseline-D) represent the background levels, while curves B, C and D contain the target signal. The full width at half maximum (FWHM) of the target feature is approximately $8~\mathrm{mm}$, consistent with the physical dimensions of the tag.

The signal-to-background ratio (SBR), which quantifies residual background modulation, is defined as
\begin{equation}
\label{deqn_eq_3}
\mathrm{SBR}=\frac{I_{\mathrm{Smax}}-I_{\mathrm{Smin}}}{I_{\mathrm{Bmax}}-I_{\mathrm{Bmin}}},
\end{equation}
where $I_{\mathrm{Smax}}$ and $I_{\mathrm{Smin}}$ are the maximum and minimum values of the signal curve (solid-line), and $I_{\mathrm{Bmax}}$ and $I_{\mathrm{Bmin}}$ are the maximum and minimum of the background curve (dashed-line). Applying Eq.~\eqref{deqn_eq_3} to the data in Fig.~\ref{fig5}a yields $\mathrm{SBR_{aC}}=0.979$ and $\mathrm{SBR_{aD}}=3.2$, indicating a more than threefold reduction of background interference after differential processing. Similar behaviour is observed at $\mathrm{X}=38~\mathrm{mm}$ (Fig.~\ref{fig5}b), where the FWHM remains $\sim 8~\mathrm{mm}$ and the SBR improves from $\mathrm{SBR_{bC}}=0.702$ to $\mathrm{SBR_{bD}}=2.77$. These results confirm that differential imaging effectively suppresses background inhomogeneity while preserving the target signal.

In radar imaging, the signal-to-noise ratio ($\mathrm{S/N}$) is defined as the ratio of signal to noise spectral densities, with larger values indicating better image quality~\cite{RN65}. Here, we take the signal spectral density as the integral of the squared electric-field strength within the target region (white box in Fig.~\ref{fig4}d) and the noise spectral density from the area outside this box. For the Rydberg probe, we obtain $\mathrm{S/N_{d}}=0.0127$, while for the metal probe (Fig.~\ref{fig4}g) we find $\mathrm{S/N_{g}}=0.0016$, corresponding to an almost eightfold improvement in S/N with the Rydberg sensor. As evident from Figs.~\ref{fig4} and \ref{fig5}, the traditional metal probe fails to resolve the target contour because of the complex interplay between probe-induced field distortion, interaction with the target and the resulting modification of the scattered field.

In summary, the Rydberg probe achieves high-resolution imaging of subwavelength targets in the reactive near-field region, with a spatial resolution of $8~\mathrm{mm}$ in the present tag-imaging experiment (corresponding to $\lambda/4$ at a wavelength of $35~\mathrm{mm}$), while independent measurements with a dual-wire resolution target establish an ultimate spatial resolution of about $0.62~\mathrm{mm}$ ($\lambda/56$) for the system~(Supplementary Information Section S.4). More importantly, its weak back-action on the field uniquely enables direct characterization of field perturbations caused by external objects. The same non-invasive methodology can therefore be used to quantify the interference introduced by conventional metal probes during near-field measurements. Such quantitative assessment of probe-induced distortions provides a foundation for developing and validating compensation algorithms, thereby advancing high-precision microwave near-field diagnostics.

\section{Conclusion}\label{sec4}

This work establishes Rydberg-atom vapour cells as practical, non-invasive probes for microwave reactive near-field imaging, bridging quantum sensing and classical antenna metrology. Using an all-dielectric, fibre-integrated head, we directly resolve the rich subwavelength structure of horn and chip-antenna fields, recover near-field patterns with almost unit structural similarity to full-wave simulations, and image scattering from deeply subwavelength metallic tags while a similarly sized metal probe fails. Together, these results show that quantum-enabled field probes can deliver accurate amplitude maps in regimes where even compact metallic antennas substantially reshape the fields they are intended to measure.

The achieved deep-subwavelength performance is enabled by a combination of spectroscopic and electromagnetic advantages and points toward a broader application space. Velocity-selective EIT--AT spectroscopy provides an SI-traceable and nearly linear transfer function between AT splitting and local microwave field amplitude, enabling faithful reconstruction of steep near-field gradients across both reactive and radiating regions, while full-wave RCS calculations confirm that the quartz-cell probe scatters substantially less than comparably sized metal antennas over all viewing angles, consistent with the experimentally observed non-invasiveness (Supplementary Information Section S.5). Combined with the dual-wire resolution benchmark, which demonstrates an ultimate spatial resolution of about $0.62~\mathrm{mm}$ ($\lambda/56$), this suggests that Rydberg-atom probes can access near-field information that is effectively hidden from conventional metal sensors and motivates future work on faster data acquisition, further miniaturization and on-chip photonic integration~\cite{RN66}, and vector- and phase-resolved measurements for antenna metrology, integrated-circuit testing and compact device characterization.

\section*{Acknowledgment}

We acknowledge support from the National Key R\&D Program of China (grant no.~2022YFA1404003), the National Natural Science Foundation of China (grants T2495252, 12104279, 123B2062, 12574318, 62201481), the Innovation Program for Quantum Science and Technology (Grant No.~2021ZD0302100), the Fund for Shanxi `1331 Project' Key Subjects Construction and the Bairen Project of Shanxi Province.

\section*{Author contributions}
L.Z. and M.J. conceived and developed the project. C.H. and M.J. performed the experiments and analyzed the data. M.J., L.Z. and H.Z. designed and constructed the experimental setup. C.H., M.J. and Z.L. wrote the manuscript. All authors discussed the results and contributed to revising the manuscript.

\section*{Data availability}
The data that support the findings of this study are available from the corresponding author upon reasonable request. 

\section*{Conflict of interest} 
The authors declare no competing interests.

\bibliography{ref}
\end{document}



\bibcite{S1}{{S1}{2001}{Lacomme et~al.}{}}
\bibcite{S2}{{S2}{2008}{Vinetti}{}}
\bibcite{S3}{{S3}{2004}{Wang et~al.}{}}

\makeatletter
\renewcommand{\bibnumfmt}[1]{[#1]}
\makeatother
\appendix

\renewcommand{\thesection}{}
\section*{Supplementary Information}
\addcontentsline{toc}{section}{Supplementary Information}
\setcounter{figure}{0}
\setcounter{table}{0}
\setcounter{equation}{0}
\renewcommand{\thefigure}{S\arabic{figure}}
\renewcommand{\thetable}{S\arabic{table}}
\renewcommand{\theequation}{S.\arabic{equation}}
\renewcommand{\thesubsection}{S.\arabic{subsection}}
\subsection{Velocity-selective EIT scheme in the non-invasive single-ended fibre-integrated Rydberg probe}
\label{Supplementary_subsec1}

\begin{figure}[!htbp]
\centering
\includegraphics[width=0.9\textwidth]{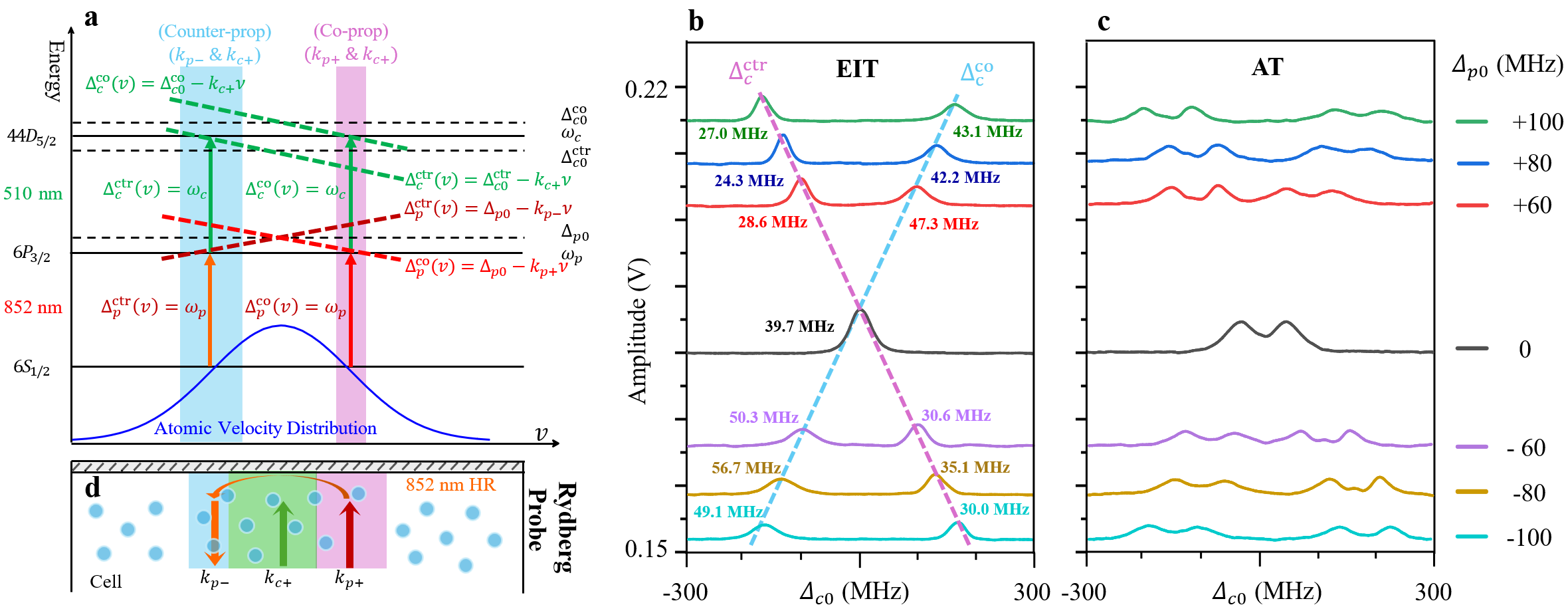}
\caption{\textbf{Velocity-selective EIT--Autler--Townes spectroscopy in the non-invasive single-ended fibre-integrated Rydberg near-field probe.}
\textbf{a} Velocity-selection scheme in the single-ended geometry. The \(510~\mathrm{nm}\) coupling beam propagates along \(+z\) (\(k_{c+}\)), while the \(852~\mathrm{nm}\) probe is retroreflected by the HR coating and forms co- and counter-propagating components (\(k_{p+}\) and \(k_{p-}\)). For a fixed probe detuning \(\Delta_{p0}\) (defined with respect to the zero-velocity transition), the two probe components address two opposite velocity classes, giving two EIT branches at different scanned coupling detunings \(\Delta_{c0}\).
\textbf{b} Measured EIT spectra as a function of the scanned coupling detuning \(\Delta_{c0}\) for different fixed probe detunings \(\Delta_{p0}\) (color-coded, in MHz). At \(\Delta_{p0}=0\) the co- and counter-propagating branches overlap near \(\Delta_{c0}=0\); for \(\Delta_{p0}\neq 0\) the two peaks separate approximately linearly and symmetrically, \(\Delta_{c0}^{\mathrm{co/ctr}}\approx \pm (k_c/k_p)\Delta_{p0}\). Numbers next to the peaks denote the full width at half maximum.
\textbf{c} Corresponding Autler--Townes spectra in the presence of the microwave field, recorded under the same conditions and plotted versus \(\Delta_{c0}\). The AT splitting \(\Delta f\) of the narrow counter-propagating EIT branch (e.g., at \(\Delta_{p0}=+80~\mathrm{MHz}\)) is used to extract the local microwave electric-field amplitude via Eq.~\eqref{deqn_eq_12}.
\textbf{d} Schematic of the single-ended probe head showing the three optical wavevectors in the vapour cell (\(k_{p-}\), \(k_{c+}\), \(k_{p+}\)) generated by retroreflection of the probe.}
\label{figS1}
\end{figure}

The non-invasive probe measures the local microwave electric field via electromagnetically induced transparency and Autler--Townes (EIT--AT) splitting in a four-level cesium ladder. As shown in Fig.~\ref{figS1}a, a probe laser at \(852~\mathrm{nm}\) drives \(\lvert 6S_{1/2} \rangle \rightarrow \lvert 6P_{3/2} \rangle\), a coupling laser at \(510~\mathrm{nm}\) drives \(\lvert 6P_{3/2} \rangle \rightarrow \lvert 44D_{5/2} \rangle\), and a microwave field at \(8.556~\mathrm{GHz}\) couples \(\lvert 44D_{5/2} \rangle \rightarrow \lvert 45P_{3/2} \rangle\). In the presence of the microwave field, the EIT resonance splits into two Autler--Townes components with separation \(\Delta f\), yielding
\begin{equation}
\label{deqn_eq_12}
|E| = \frac{h}{\wp_{\mathrm{RF}}}\,\Delta f,
\end{equation}
where \(h\) is Planck's constant and \(\wp_{\mathrm{RF}}\) is the transition dipole moment.

In the single-ended probe, the \(852~\mathrm{nm}\) beam is retroreflected by the HR coating, producing co- and counter-propagating probe components with respect to the \(510~\mathrm{nm}\) coupling beam (Fig.~\ref{figS1}d). As a result, the measured EIT--AT spectra contain two Doppler-selected branches (co/ctr), unlike the conventional geometry with only counter-propagating probe and coupling beams. Below we use a one-dimensional Doppler model to account for the branch positions, relative amplitudes and linewidths in Fig.~\ref{figS1}b,c.

We choose the \(z\)-axis along the propagation direction of the \(510~\mathrm{nm}\) coupling beam and denote by \(v\) the atomic velocity component along \(z\). In the experiment, the probe- and coupling-laser detunings are defined with respect to the zero-velocity transitions and are denoted by \(\Delta_{p0}\) (set value) and \(\Delta_{c0}\) (scan variable), respectively. Because the coupling beam propagates along \(+z\) with \(k_{c+}=+k_c=\omega_c/c\), the Doppler-shifted coupling detuning is
\begin{equation}
\Delta_c(v)=\Delta_{c0}-k_{c+}v.
\end{equation}
The retroreflected \(852~\mathrm{nm}\) probe provides two components with \(k_{p+}=+k_p=\omega_p/c\) (co-propagating) and \(k_{p-}=-k_p\) (counter-propagating), yielding
\begin{align}
\Delta_p^{\mathrm{co}}(v)  &= \Delta_{p0}-k_{p+}v,\\
\Delta_p^{\mathrm{ctr}}(v) &= \Delta_{p0}-k_{p-}v=\Delta_{p0}+k_p v.
\end{align}

For each branch, the two-photon detuning in the atomic frame is
\begin{equation}
\delta_{2\gamma}^{\mathrm{co/ctr}}(v)
= \Delta_p^{\mathrm{co/ctr}}(v)+\Delta_c(v)
= (\Delta_{p0}+\Delta_{c0})-K_{\mathrm{co/ctr}}v,
\label{eq:d2g}
\end{equation}
with
\begin{equation}
K_{\mathrm{co}}=k_{p+}+k_{c+}=k_p+k_c,
\qquad
K_{\mathrm{ctr}}=k_{p-}+k_{c+}=k_c-k_p.
\end{equation}

For a given \(\Delta_{p0}\), the EIT signal in each branch is dominated by atoms close to one-photon probe resonance and two-photon resonance, which we summarize as
\begin{equation}
|\Delta_p^{\mathrm{co/ctr}}(v)| \lesssim \gamma_p,
\qquad
|\delta_{2\gamma}^{\mathrm{co/ctr}}(v)| \lesssim \gamma_{2\gamma},
\label{eq:conds}
\end{equation}
where \(\gamma_p\) and \(\gamma_{2\gamma}\) are the effective one- and two-photon linewidths.

Imposing \(\Delta_p^{\mathrm{co/ctr}}(v)\approx 0\) gives the probe-selected velocity classes
\begin{equation}
v_{\mathrm{co/ctr}}\approx \pm \frac{\Delta_{p0}}{k_p},
\end{equation}
where the upper (lower) sign corresponds to the co-propagating (counter-propagating) probe component. Evaluating Eq.~\eqref{eq:d2g} at \(v_{\mathrm{co/ctr}}\) and requiring \(\delta_{2\gamma}^{\mathrm{co/ctr}}(v_{\mathrm{co/ctr}})\approx 0\) yields the positions of the two EIT branches on the scanned \(\Delta_{c0}\) axis,
\begin{equation}
\Delta_{c0}^{\mathrm{co/ctr}} \approx \pm \frac{k_c}{k_p}\,\Delta_{p0},
\label{eq:Dc_summary}
\end{equation}
consistent with the symmetric splitting observed in Fig.~\ref{figS1}b.

The relative \emph{heights} of the two branches follow from the Doppler acceptance in velocity space. From Eq.~\eqref{eq:d2g}, the two-photon condition \(|\delta_{2\gamma}^{\mathrm{co/ctr}}(v)|\lesssim\gamma_{2\gamma}\) corresponds to a characteristic velocity width
\begin{equation}
\Delta v_{\mathrm{co/ctr}} \sim \frac{\gamma_{2\gamma}}{|K_{\mathrm{co/ctr}}|}.
\end{equation}
For the \(852~\mathrm{nm}/510~\mathrm{nm}\) ladder, \(|k_c-k_p|\ll|k_c+k_p|\), hence \(|K_{\mathrm{ctr}}|\ll|K_{\mathrm{co}}|\) and \(\Delta v_{\mathrm{ctr}}\gg\Delta v_{\mathrm{co}}\). The counter-propagating branch therefore integrates over a much broader velocity slice of the Maxwell--Boltzmann distribution and appears stronger, whereas the co-propagating branch is weaker.

The \emph{linewidth} versus the scanned coupling detuning \(\Delta_{c0}\) can be understood from the resonant velocity \(v_{\mathrm{res}}^{\mathrm{co/ctr}}(\Delta_{c0})\) defined by \(\delta_{2\gamma}^{\mathrm{co/ctr}}(v_{\mathrm{res}})\approx 0\), i.e.
\begin{equation}
v_{\mathrm{res}}^{\mathrm{co/ctr}}(\Delta_{c0}) \simeq \frac{\Delta_{p0}+\Delta_{c0}}{K_{\mathrm{co/ctr}}},
\qquad
\frac{\mathrm{d}v_{\mathrm{res}}^{\mathrm{co/ctr}}}{\mathrm{d}\Delta_{c0}} \simeq \frac{1}{K_{\mathrm{co/ctr}}}.
\end{equation}
In the counter-propagating branch, \(|K_{\mathrm{ctr}}|\) is small, so a modest change in \(\Delta_{c0}\) corresponds to a large shift in \(v_{\mathrm{res}}\), rapidly moving the resonant atoms into the tails of the velocity distribution and yielding a narrow spectral feature. In contrast, the co-propagating branch has much larger \(|K_{\mathrm{co}}|\), so \(v_{\mathrm{res}}\) varies slowly with \(\Delta_{c0}\), producing a broader but weaker feature.

This Doppler picture explains why, for \(\Delta_p\neq 0\), the EIT spectra split into a weak, broad co-propagating branch and a strong, narrow counter-propagating branch (Fig.~\ref{figS1}b). In the main experiments we choose \(\Delta_p=+80~\mathrm{MHz}\) to spectrally isolate the counter-propagating branch and use its Autler--Townes splitting to extract the local microwave electric field via Eq.~\eqref{deqn_eq_12}.

\subsection{Quantifying fidelity of microwave imaging using the structural similarity index (SSIM)}
\label{Supplementary_subsec2}

To complement the brief definition in Eq.~(2) of the main text, we summarize here the standard decomposition of the structural similarity index (SSIM)~\cite{S3} into luminance, contrast and structure components for normalized microwave electric-field maps.

The luminance components $\mu_i$ and $\mu_j$, representing the local mean field intensity in images $i$ and $j$, are computed from the normalized electric-field strengths $E_{i,k}$ and $E_{j,k}$ within a local window:
\begin{equation}
\label{deqn_eq_4}
\mu_{i} = \frac{1}{N} \sum_{k=1}^{N} E_{i,k},
\end{equation}
where $N$ is the number of pixels in the window. The contrast components $\sigma_i$ and $\sigma_j$, which quantify local variations in the field distribution, are given by the standard deviations
\begin{equation}
\label{deqn_eq_5}
\sigma_i = \left( \frac{1}{N-1} \sum_{k=1}^N (E_{i,k} - \mu_i)^2 \right)^{1/2},
\end{equation}
and the structural component $\sigma_{ij}$, capturing the covariance between the two field maps, is
\begin{equation}
\label{deqn_eq_6}
\sigma_{ij} = \frac{1}{N-1} \sum_{k=1}^N (E_{i,k} - \mu_i)(E_{j,k} - \mu_j).
\end{equation}

Using these quantities, the luminance comparison function is
\begin{equation}
\label{deqn_eq_7}
l(i,j) = \frac{2\mu_i\mu_j + C_1}{\mu_i^2 + \mu_j^2 + C_1},
\end{equation}
the contrast comparison function is
\begin{equation}
\label{deqn_eq_8}
c(i,j) = \frac{2\sigma_i\sigma_j + C_2}{\sigma_i^2 + \sigma_j^2 + C_2},
\end{equation}
and the structure comparison function, related to the correlation coefficient between the two normalized field distributions, is
\begin{equation}
\label{deqn_eq_9}
s(i,j) = \frac{\sigma_{ij} + C_3}{\sigma_i \sigma_j + C_3}.
\end{equation}
The general form of the SSIM index is then expressed as
\begin{equation}
\label{deqn_eq_10}
\mathrm{SSIM}(i,j) = [l(i,j)]^{\alpha} [c(i,j)]^{\beta} [s(i,j)]^{\gamma},
\end{equation}
where $\alpha > 0$, $\beta > 0$ and $\gamma > 0$ weight the relative contributions of luminance, contrast and structure.

For simplicity, we set $\alpha = \beta = \gamma = 1$ and $C_3 = C_2/2$, and use $C_1=(K_1L)^2$ and $C_2=(K_2L)^2$ with the standard values $K_1=0.01$, $K_2=0.03$ and $L=1$ the dynamic range of the normalized field. Under these choices, substituting Eqs.~\eqref{deqn_eq_7}–\eqref{deqn_eq_9} into Eq.~\eqref{deqn_eq_10} recovers the SSIM expression given in Eq.~(2) of the main text, which takes values between $0$ and $1$ and increases with the similarity between two microwave field maps.

\subsection{Microwave non-invasive imaging in the compact near field of a chip antenna}
\label{Supplementary_subsec3}

The compact near-field (CNF) microwave electric field—defined here as the reactive-near-field region of an electrically small chip antenna fully contained within the probe footprint—is characterized using the Rydberg probe. CNF regions of this type are common in space-constrained environments, such as patch or chip antennas in MIMO systems~\cite{S2}. By reciprocity, the far-field radiation pattern of a transmitting antenna is identical to its reception pattern, so accurate reconstruction of the reception pattern from CNF measurements requires a highly non-invasive probe.

\begin{figure}[!htbp]
\centering
\includegraphics[width=1\textwidth]{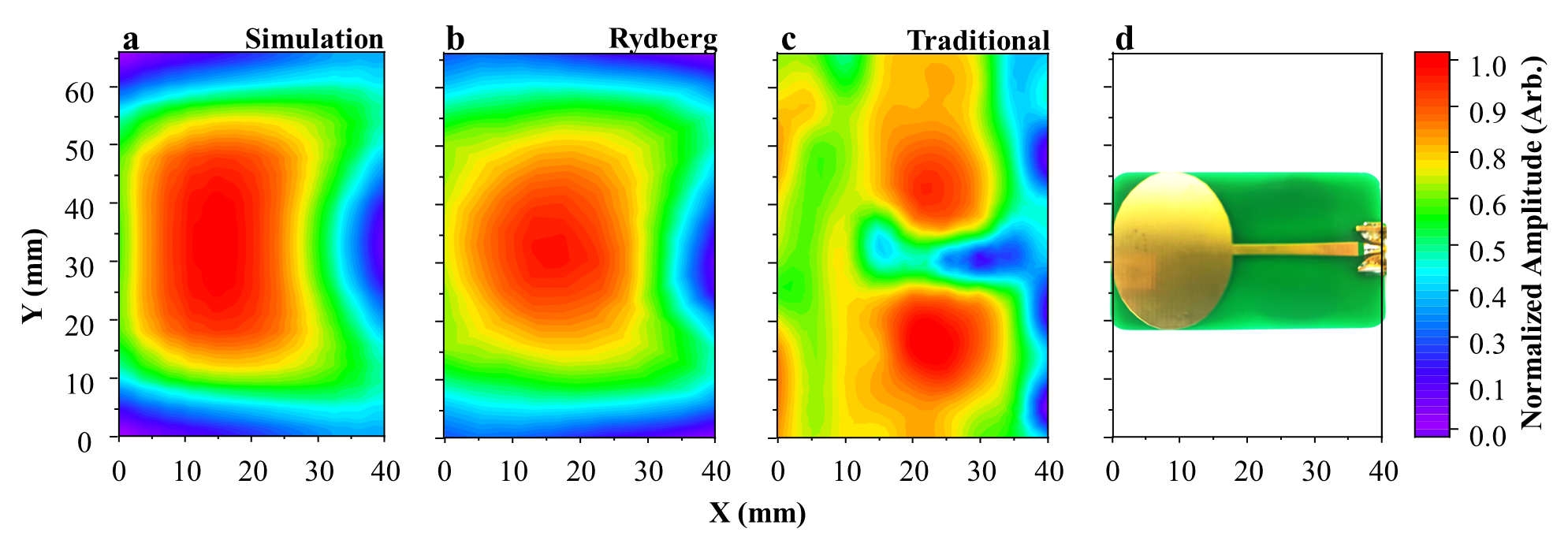}
\caption{\textbf{Microwave electric-field distributions in the compact near-field region of a chip antenna.}
\textbf{a} CST-simulated field distribution used as a non-invasive reference.
\textbf{b} Measured field distribution obtained with the Rydberg-atom probe.
\textbf{c} Field distribution acquired using a traditional metal antenna of the same model as the AUT.
\textbf{d} Physical dimensions and mounting geometry of the UWB chip antenna (XY plane, $\mathrm{Z} = 15~\mathrm{mm}$, $\Delta \mathrm{X} = 5~\mathrm{mm}$, $\Delta \mathrm{Y} = 5~\mathrm{mm}$).}
\label{figS2}
\end{figure}

In our implementation, the AUT is a commercial ultra-wideband (UWB) chip antenna with an approximately elliptical radiation profile (see Fig.1b and Fig.~\ref{figS2}d). Using Eq.~(1) in the main text, the corresponding compact near-field (CNF) region spans \(0\) to \(35~\mathrm{mm}\) at an operating frequency of \(8.556~\mathrm{GHz}\).

We characterize the CNF field distribution using three approaches: CST simulation as a nominal non-invasive reference (Fig.~\ref{figS2}a), non-invasive imaging with the Rydberg probe (Fig.~\ref{figS2}b), and, for comparison, imaging with a conventional metallic probe antenna of the same model as the AUT (Fig.~\ref{figS2}c). The antenna geometry and mounting configuration are shown in Fig.~\ref{figS2}d.

The Rydberg-based measurement in Fig.~\ref{figS2}b faithfully reproduces the shape and extent of the simulated elliptical radiation region. In contrast, when a second, identical UWB chip antenna is used as a probe (Fig.~\ref{figS2}c), the measured field pattern is strongly distorted. This distortion is primarily caused by strong mutual coupling at close range: the metallic probe antenna significantly perturbs the CNF that it is intended to measure, leading to a distorted field pattern.

Quantitative comparison using SSIM gives $\mathrm{SSIM}(b,a) = 0.918$ for the Rydberg probe and $\mathrm{SSIM}(c,a) = 0.297$ for the metal-antenna probe, indicating that the Rydberg measurement is in much closer agreement with the simulated reference. These results highlight the importance of a genuinely non-invasive probe for compact near-field characterization in integrated and densely packed antenna systems.

\subsection{Spatial resolution of non-invasive microwave reactive-near-field imaging}
\label{Supplementary_subsec4}

\begin{figure}[!htbp]
\centering
\includegraphics[width=1\textwidth]{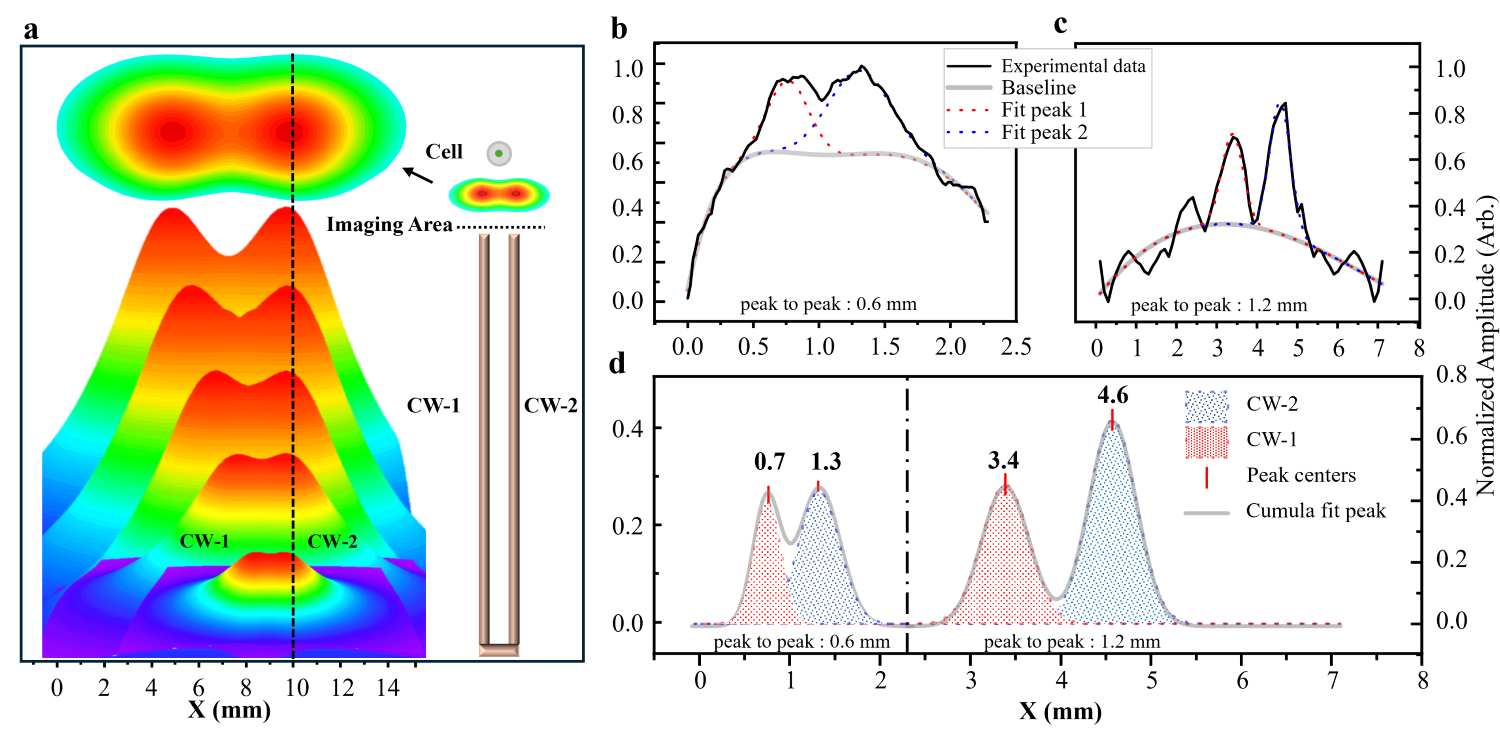}
\caption{\textbf{Spatial-resolution characterization of non-invasive microwave near-field imaging.}
\textbf{a} CST-simulated electric-field magnitude near the wire tips for \(\Delta d = 1\)–\(5~\mathrm{mm}\) (step \(1~\mathrm{mm}\)), showing that the peak-to-peak spacing tracks \(\Delta d\).
\textbf{b} One-dimensional electric-field profile along $\mathrm{X}$ for a wire separation $\Delta d = 1.2~\mathrm{mm}$ ($\Delta \mathrm{X} = 0.1~\mathrm{mm}$).
\textbf{c} Corresponding profile for $\Delta d = 0.6~\mathrm{mm}$ ($\Delta \mathrm{X} = 0.03~\mathrm{mm}$).
\textbf{d} Processed doublet curves from panels \textbf{b} and \textbf{c} after baseline subtraction and peak fitting.}
\label{figS3}
\end{figure}

To evaluate the spatial resolution of the Rydberg-atom probe for microwave reactive-near-field (RNF) imaging, we employ a dual--copper--wire target, following the standard radar-imaging definition of spatial resolution~\cite{S1}: the minimum separation at which two point targets can be distinguished. This configuration effectively provides a calibrated pair of point-like scatterers for resolution testing.

As shown in Fig.~\ref{figS3}a, two parallel copper wires (diameter \(0.6~\mathrm{mm}\)) are driven by identical \(8.556~\mathrm{GHz}\) microwave signals with equal power. CST simulations for \(\Delta d = 1\)–\(5~\mathrm{mm}\) show two localized maxima near the wire tips whose peak-to-peak separation follows the imposed wire spacing \(\Delta d\) (each map is normalized independently for clarity). This validates the use of the measured tip-field peak separation as a quantitative metric for calibrating the spatial resolution of the RNF imaging system.

Figs.~\ref{figS3}b and \ref{figS3}c show the measured one-dimensional profiles for wire spacings of \(\Delta d = 1.2~\mathrm{mm}\) and \(\Delta d = 0.6~\mathrm{mm}\), respectively. The $1.2~\mathrm{mm}$ spacing yields a well-resolved doublet, with two distinct peaks that can be individually associated with each wire. The $0.6~\mathrm{mm}$ spacing probes the near-limit of the system’s resolution, where the two peaks begin to overlap yet remain marginally distinguishable.

To enhance the visibility of the peaks, the raw data in Figs.~\ref{figS3}b and \ref{figS3}c are processed using a combined baseline-subtraction and peak-fitting procedure. Regions outside the double-peak area are used to estimate the baseline, which is fitted simultaneously with the peaks. The processed curves in Fig.~\ref{figS3}d reveal two clearly separated maxima in both spacing configurations. Gaussian fits yield full widths at half maximum (FWHM) of $0.625~\mathrm{mm}$ and $0.620~\mathrm{mm}$ for the two peaks in the $1.2~\mathrm{mm}$-spacing case, closely matching the copper-wire diameter. For the $0.6~\mathrm{mm}$ spacing, the peak separation extracted from the processed data is approximately $0.62~\mathrm{mm}$, in good agreement with the actual spacing and corresponding to about $\lambda/56$ at a wavelength of $35~\mathrm{mm}$.

These measurements therefore set the experimental spatial resolution of the present RNF imaging system at approximately $0.62~\mathrm{mm}$ ($\lambda/56$).

\subsection{Radar cross-section of the Rydberg-atom probe}
\label{Supplementary_subsec5}

When an electromagnetic (EM) wave impinges on a target, part of the energy is reradiated in different directions; this reradiation is termed scattering, and the target is described as a scatterer. The radar cross-section (RCS) quantifies the effective area that scatters EM power back toward the source or receiver~\cite{S2}. For microwave near-field probes, the RCS is a key figure of merit: a smaller RCS implies weaker perturbation of the ambient field and thus more non-invasive imaging.

\begin{figure}[!htbp]
\centering
\includegraphics[width=0.9\textwidth]{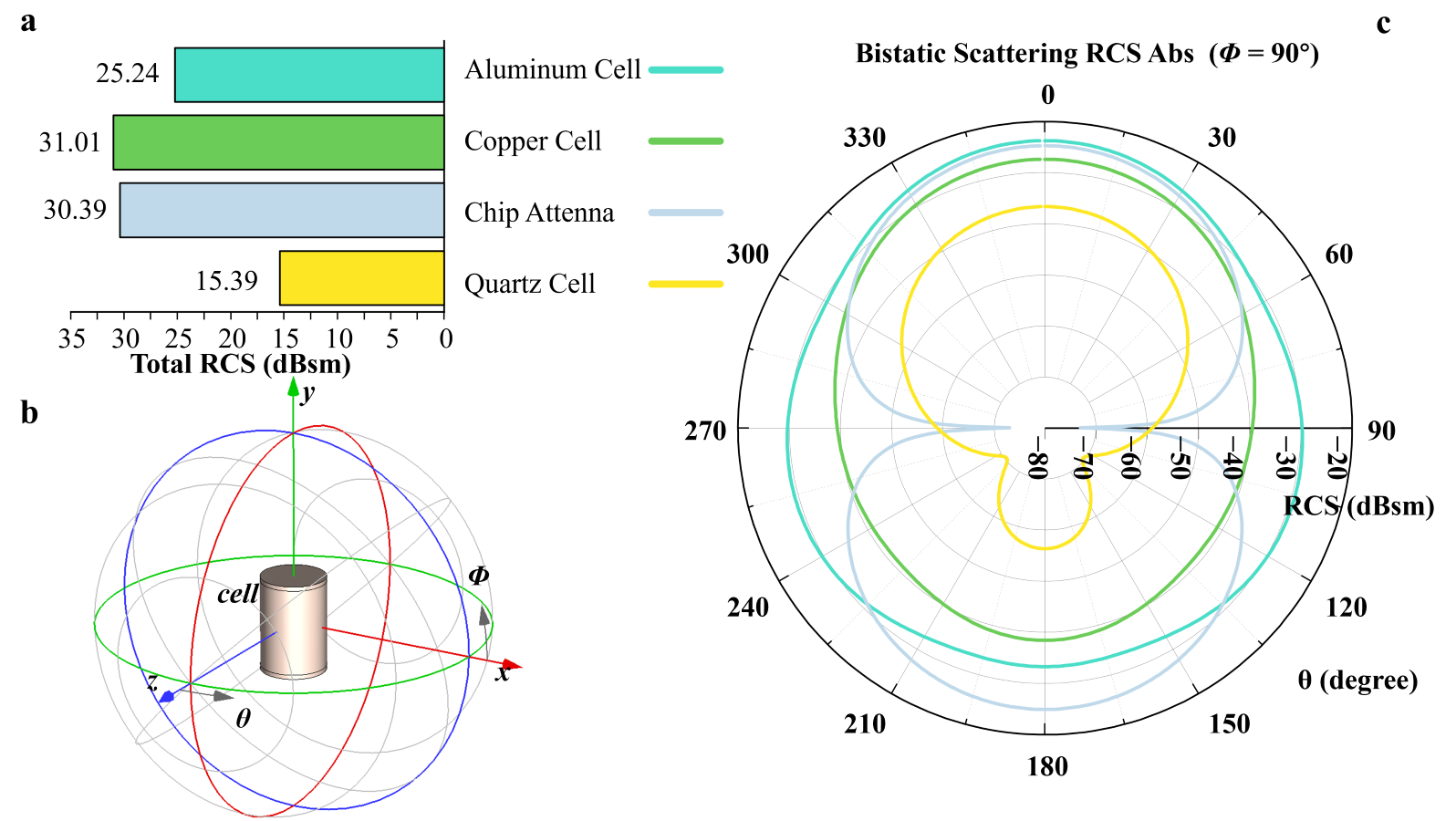}
\caption{\textbf{Radar cross-section (RCS) simulations of the Rydberg-atom probe.}
\textbf{a} Total RCS of a chip antenna and macroscopic cells of different materials. Each cell is a cylinder with diameter $15~\mathrm{mm}$ and height $20~\mathrm{mm}$.
\textbf{b} Schematic of the target position and microwave polarization direction.
\textbf{c} RCS of microscale cells of different materials when $\phi = 90^\circ$.}
\label{figS4}
\end{figure}

We therefore use CST simulations to compare the RCS of various candidate near-field probes under conditions that match the experimental geometry. In all simulations, the microwave frequency, polarization and incidence direction are chosen as in the experiment. As illustrated in Fig.~\ref{figS4}b, where the coordinate axes and the angles $\phi$ and $\theta$ are defined based on the probe geometry, the incident wave propagates along the positive $\mathrm{Z}$-direction, with its polarization along $\phi = 0^\circ$. The simulated targets include a quartz vapour cell, metal cells of various materials and a UWB chip antenna.

The vapour cell and chip-antenna geometries are chosen to match those used in the experiment. The Rydberg-atom vapour cell is made of quartz, whereas traditional metal antennas are typically fabricated from copper or aluminium. Figure~\ref{figS4}a compares the total RCS for the chip antenna and for macroscopic cylindrical cells of different materials; Fig.~\ref{figS4}c shows the corresponding results for scaled-down cells. Among all structures tested, the quartz cell exhibits the lowest total RCS over all polar angles $\theta$. For cells of identical size and shape, the quartz cell has a substantially lower RCS than copper or aluminium cells. Despite its somewhat different shape and slightly smaller volume, the chip antenna still shows a higher RCS than the quartz cell across the angular range.

An additional advantage of the Rydberg probe is that its size is not constrained by the microwave wavelength: the vapour-cell dimensions can be reduced far below $\lambda$ without compromising sensing performance, whereas conventional metal antennas become inefficient when miniaturized into the deep subwavelength regime. Taken together, these simulations show that the Rydberg-atom microwave electric-field sensor has a pronounced advantage over traditional metallic antennas for non-invasive near-field imaging, owing to its intrinsically low RCS and scalable, all-dielectric geometry.
